\documentclass[12pt]{iopart}

\usepackage{iopams}  
\usepackage{bm}
\usepackage{graphicx}
\usepackage[centerlast]{subfigure}

\newcommand{\St}{{\mbox{\it St}}}
\newcommand{\Ku}{{\mbox{\it Ku}}}

\begin{document}

\title[A phenomenological approach to the clustering of heavy
  particles in turbulent flows]{Toward a phenomenological approach to
  the clustering of heavy particles in turbulent flows}

\author{J\'er\'emie Bec}
\address{CNRS UMR6202, Laboratoire Cassiop\'ee, OCA, BP4229, 06304
  Nice Cedex 4, France}
\ead{jeremie.bec@obs-nice.fr}

\author{Rapha\"el Ch\'etrite} \address{Laboratoire de Physique, ENS
Lyon, 11 All\'ee d'Italie, 69007 Lyon, France}

\begin{abstract}
A simple model accounting for the ejection of heavy particles from the
vortical structures of a turbulent flow is introduced. This model
involves a space and time discretization of the dynamics and depends
on only two parameters: the fraction of space-time occupied by
rotating structures of the carrier flow and the rate at which
particles are ejected from them. The latter can be heuristically
related to the response time of the particles and hence measure their
inertia. It is shown that such a model reproduces qualitatively most
aspects of the spatial distribution of heavy particles transported by
realistic flows. In particular the probability density function of the
mass $m$ in a cell displays an power-law behavior at small values and
decreases faster than exponentially at large values. The dependence of
the exponent of the first tail upon the parameters of the dynamics is
explicitly derived for the model. The right tail is shown to decrease
as $\exp (-C m \log m)$. Finally, the distribution of mass averaged
over several cells is shown to obey rescaling properties as a function
of the coarse-grain size and of the ejection rate of the
particles. Contrarily to what has been observed in direct numerical
simulations of turbulent flows (Bec \textit{et al.}, nlin.CD/0608045),
such rescaling properties are only due in the model to the mass
dynamics of the particles and do not involve any scaling properties in
the spatial structure of the carrier flow.

\end{abstract}

\maketitle

\section{Introduction}

Understanding the dynamics of small-size tracer particles or of a
passive field transported by an incompressible turbulent flow plays an
important role in the description of several natural and industrial
phenomena.  For instance it is well known that turbulence has the
property to induce an efficient mixing over the whole range of length
and time scales spanned by the turbulent cascade of kinetic energy
(see e.g.~\cite{d05}). Describing quantitatively such a mixing has
consequences in the design of engines, in the prediction of pollutant
dispersion or in the development of climate models accounting for
transport of salinity and temperature by large-scale ocean streams.

However, in some settings, the suspended particles have a finite size
and a mass density very different from that of the fluid. Thus they
can hardly be modeled by tracers because they have inertia.  In order
to fully describe the dynamics of such inertial particles, one has to
consider many forces that are exerted by the fluid even in the simple
approximation where the particle is a hard sphere much smaller than
the smallest active scale of the fluid flow \cite{mr83}. Nevertheless
the dynamics drastically simplifies in the asymptotics of particles
much heavier than the carrier fluid. In that case, and when buoyancy
is neglected, they interact with the flow only through a viscous drag,
so that their trajectories are solutions to the Newton equation\,:
\begin{equation}
\frac{d^2\bm X}{dt^2} = -\frac{1}{\tau} \left[ \frac{d\bm X}{dt} -
  \bm u(\bm X,t)\right]\,,
\end{equation}
where $\bm u$ denotes the underlying fluid velocity field and $\tau$
is the response time of the particles. Even if the carrier fluid is
incompressible, the dynamics of such heavy particles lags behind that
of the fluid and is not volume-preserving. At large times particles
concentrate on singular sets evolving with the fluid motion, leading
to the appearance of strong spatial inhomogeneities dubbed
preferential concentrations. At the experimental level such
inhomogeneities have been known for a long time (see~\cite{ef94} for a
review). At present the statistical description of particle
concentration is a largely open question with many applications. We
mention the formation of rain droplets in warm clouds~\cite{ffs02},
the coexistence of plankton species~\cite{lp01}, the dispersion in the
atmosphere of spores, dust, pollen, and chemicals~\cite{s86}, and the
formation of planets by accretion of dust in gaseous circumstellar
disks~\cite{pl01}.

The dynamics of inertial particles in turbulent flows involves a
competition between two effects: on the one hand particles have a
dissipative dynamics, leading to the convergence of their trajectories
onto a dynamical attractor~\cite{b05}, and on the other hand, the
particles are ejected from the coherent vortical structures of the
flow by centrifugal inertial forces~\cite{m87}. The simultaneous
presence of these two mechanisms has so far led to the failure of all
attempts made to obtain analytically the dynamical properties or the
mass distribution of inertial particles.  In order to circumvent such
difficulties a simple idea is to tackle independently the two aspects
by studying toy models, either for the fluid velocity field, or for
the particle dynamics that are hopefully relevant in some asymptotics
(small or large response times, large observation scales,
etc.). Recently an important effort has been made in the understanding
of the dynamics of particles suspended in flows that are
$\delta$-correlated in time, as in the case of the well-known
Kraichnan model for passive tracers~\cite{k68}. Such settings, which
describe well the limit of large response time of the particles,
allows one to obtain closed equations for density correlations by
Markov techniques. The $\delta$-correlation in time, of course, rules
out the presence of any persistent structure in the flow; hence any
observed concentrations can only stem from the dissipative
dynamics. Most studies in such simplified flows dealt with the study
of the separation between two particles~\cite{p02,mw04,dmow05,
detal06,bch06}.

Recent numerical studies in fully developped turbulent
flows~\cite{betal06} showed that the spatial distribution of particles
at lengthscales within the inertial range are strongly influenced by
the presence of voids at all active scales spanned by the turbulent
cascade of kinetic energy.  The presence of these voids has a
noticeable statistical signature on the probability density function
(PDF) of the coarse-grained mass of particles which displays an
algebraic tail at small values.  To understand at least from a
qualitative and phenomenological viewpoint such phenomena, it is
clearly important to consider flows with persistent vortical
structures which are ejecting heavy particles.  For this purpose, we
introduce in this paper a toy model where the vorticity field of the
carrier flow is assumed piecewise constant in both time and space and
takes either a finite fixed value $\omega$ or vanishes. In addition to
this crude simplification of the spatial structure of the fluid
velocity field we assume that the particle mass dynamics obeys the
following rule: during each time step there is a mass transfer between
the cells having vorticity $\omega$ toward the neighboring cells where
the vorticity vanishes. The amount of mass escaping to neighbors is at
most a fixed fraction $\gamma$ of the mass initially contained in the
ejecting cell. We show that such a simplified dynamics reproduces many
aspects of the mass distribution of heavy particles in incompressible
flow. In particular, we show that the PDF of the mass of inertial
particles has an algebraic tail at small values and decreases as
$\exp(-A\,m\,\log m)$ when $m$ is large.  Analytical predictions are
confirmed by numerical experiments in one and two dimensions.

In section~\ref{sec:ejection} we give some heuristic motivations for
considering such a model and a qualitative relation between the
ejection rate $\gamma$ and the response time $\tau$ of the heavy
particles.  Section \ref{sec:model} consists in a precise definition
of the model in one dimension and in its extension to higher
dimensions. Section \ref{sec:pdfm} is devoted to the study in the
statistical steady state of the PDF of the mass in a single cell. In
section~\ref{sec:pdfcoarse} we study the mass distribution averaged
over several cells to gain some insight on the scaling properties in
the mass distribution induced by the model.
Section~\ref{sec:conclusion} encompasses concluding remarks and
discussions on the extensions and improvements of the model that are
required to get a more quantitative insight on the preferential
concentration of heavy particles in turbulent flows.

\section{Ejection of heavy particles from eddies}
\label{sec:ejection}

\begin{figure}[ht]
  \centerline{\includegraphics[width=0.666\textwidth]{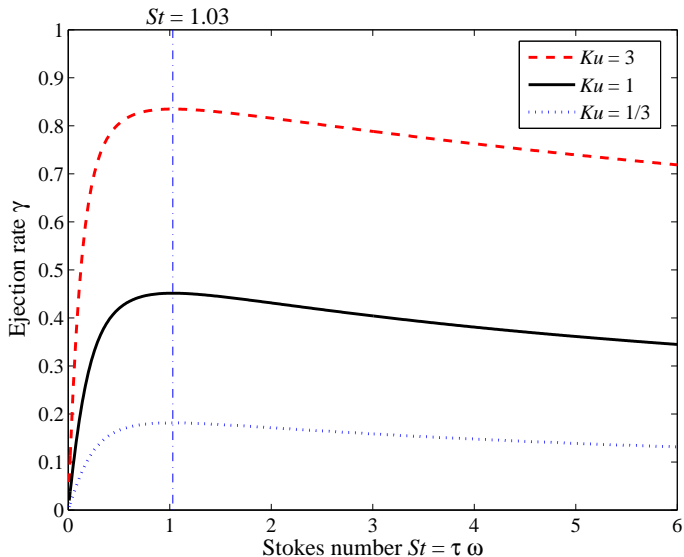}}
  \vspace{-10pt}
  \caption{Fraction of the mass of particles that are uniformly
  distributed and ejected from an eddy of arbitrary size $\ell$ as a
  function of the Stokes number $\St=\tau\omega$.  The various curves
  refer to different values of the Kubo number $\Ku$ as labeled. }
  \label{fig:gamma_fn_St}
\end{figure}

The goal of this section is to give some phenomenological arguments
explaining why the model which is shortly described above, might be of
relevance to the dynamics of heavy particles suspended in
incompressible flows. In particular we explain why a fraction of the
mass of particles exits a rotating region and give a qualitative
relation between the ejection rate $\gamma$ and the response time
$\tau$ entering the dynamics of heavy particles.  For this we focus on
the two-dimensional case and consider a cell of size $\ell$ where the
fluid vorticity $\omega$ is constant and the fluid velocity vanishes
at the center of the cell. This amounts to considering that the fluid
velocity is linear in the cell with a profile given to leading order
by the strain matrix. Having a constant vorticity in a
two-dimensional incompressible flow means that we focus on cases where
the two eigenvalues of the strain matrix are purely imaginary complex
conjugate. The particle dynamics reduces to the second-order
two-dimensional linear system
\begin{equation}
  \frac{d^2\bm X}{dt^2} = -\frac{1}{\tau}\,\frac{d\bm X}{dt} +
      \frac{\omega}{\tau} \left[\!\!\!\begin{array}{rl} 0 & 1 \\ -1 &
      0
      \end{array} \right]\, \bm X \,.
\end{equation}
It is easily checked that the four eigenvalues of the evolution matrix
are the following complex conjugate
\begin{equation}
  \lambda_{\pm,\pm} = \frac{-1 \pm \sqrt{1 \pm 4 i \tau\omega}}{2\tau}\,.
\end{equation}
Only $\lambda_{+,-}$ and $\lambda_{+,+}$ have a positive real part
which is equal to
\begin{equation}
  \mu = \frac{-1 +\frac{1}{2}
  \sqrt{2\sqrt{1+16\tau^2\omega^2}+2}}{2\tau}\,.
\end{equation}
This means that the distance of the particles to the center of the
cell increases exponentially fast in time with a rate $\mu$. If we now
consider that the particles are initially uniformly distributed inside
the cell, we obtain that the mass of particles remaining in it
decreases exponentially fast in time with a rate equal to
$-2\mu$. Namely the mass of particles which are still in the cell at
time $T$ is
\begin{equation}
  m(T) = m(0)\,(1-\gamma) = m(0)\,
  \exp\!\!\left[-\frac{T}{\tau}\!\left(-1 +\frac{1}{2}
  \sqrt{2\sqrt{1+16\tau^2\omega^2}+2} \right)\!\right]\!\!.
\end{equation}
The rate $\gamma$ at which particles are expelled from the cell
depends upon the response time $\tau$ of the particles and upon two
characteristic times associated to the fluid velocity. The first is
the time length $T$ of the ejection process which is given by the
typical life time of the structure with vorticity $\omega$.  The
second time scale is the turnover $\omega^{-1}$ which measures the
strength of the eddy. There are hence two dimensionless parameters
entering the ejection rate $\gamma$: the Stokes number $\St =
\tau\omega$ giving a measure of inertia and the Kubo number $\Ku =
T\omega$ which is the ratio between the correlation time of structures
and their eddy turnover time. One hence obtain the following estimate
of the ejection rate
\begin{equation}
  \gamma = 1 - \exp\!\!\left[-\frac{\Ku}{\St}\,\left(-1 +\frac{1}{2}
  \sqrt{2\sqrt{1+16\St^2}+2} \right)\!\right]\!\!.
  \label{eq:estimgamma}
\end{equation}
The graph of the fraction of particles ejected from the cell as a
function of the Stokes number is represented in
figure~\ref{fig:gamma_fn_St} for three different values of the Kubo
number. The function goes to zero as $\Ku\,\St$ in the limit $\St\to0$
and as $\Ku\,\St^{-1/2}$ in the limit $\St\to\infty$. It reaches a
maximum which is an indication of a maximal segregation of the
particles, for $\St\approx 1.03$ independently of the value of $\Ku$.

In three dimensions, one can extend the previous approach to obtain an
ejection rate for cells with a uniform rotation, i.e.\ a constant
vorticity $\bm\omega$. There are however two main difficulties. The
first is that in three dimensions the eigenvalues of the strain matrix
in rotating regions are not anymore purely imaginary but have a real
part given by the opposite of the rate in the stretching
direction. Such a vortex stretching has to be considered to match
observation in real flows. The second difficulty stems from the fact
that the vorticity is now a vector and has a direction, so that
ejection from the cell can be done only in the directions
perpendicular to the direction of $\bm\omega$. These two difficulties
imply that the spectrum of possible ejection rates is much broader
than in the two-dimensional case. However the rough qualitative
picture is not changed.

\section{A simple mass transport model}
\label{sec:model}

\begin{figure}[ht]
  \centerline{\includegraphics[width=0.55\textwidth]{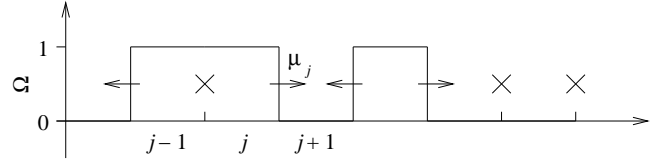}}
  \vspace{-10pt}
  \caption{Sketch of the dynamics in the one-dimensional case: the
  fluxes of mass are represented as arrows. A cross means no flux. }
  \label{fig:sketch}
\end{figure}

We here describe with details the model in one dimension and mention
at the end of the section how to generalize it to two and higher
dimensions. Let us consider a discrete partition of an interval in $N$
small cells. Each of these cell is associated to a mass which is a
continuous variable. We denote by $m_j(n)$ the mass in the $j$-th cell
at time $t=n$. At each integer time we choose randomly $N$ independent
variables; $\Omega_j=1$ with probability $p$ and $\Omega_j=0$ with
probability $1-p$. The evolution of mass between times $n$ and $n+1$
is given by:
\begin{equation}
  m_j(n+1) = \left\{ \begin{array}{ll} m_j(n) -
  \frac{\gamma}{2}\,\left[2-\Omega_{j-1}-\Omega_{j+1}\right]\,m_j(n) &
  \mbox{if } \Omega_j = 1\,, \\ m_j(n) +
  \frac{\gamma}{2}\,\left[\Omega_{j-1}\,m_{j-1}(n)+\Omega_{j+1}\,
  m_{j+1}(n)\right] & \mbox{if } \Omega_j = 0\,. \end{array} \right.
  \label{eq:massdynamics}
\end{equation}
In other terms, when $\Omega_j=1$, the $j$-th cell looses mass if
$\Omega_{j-1}=0$ or $\Omega_{j+1} = 0$, and when $\Omega_j=0$, it
gains mass if $\Omega_{j-1}=1$ or $\Omega_{j+1} = 1$. The flux of mass
between the $j$-th and the $(j+1)$-th cell is proportional $\Omega_j -
\Omega_{j+1}$ (see figure~\ref{fig:sketch}).
In particular, if $\Omega_j = \Omega_{j+1}$, no mass is transfered
between cells. When the system is supplemented by periodic boundary
conditions between the cells $N$ and 1, it is clear that the total
mass is conserved. Hereafter we assume that the mass is initially
$m_j=1$ in all cells., so that the total mass is $\sum_j m_j =
N$. Spatial homogeneity of the random process $\Omega_j$ implies that
$\langle m_j \rangle = 1$ for all later times, where the angular
brackets denote average with respect to the realizations of the
$\Omega_j$'s.

A noticeable advantage of such a model for mass transportation is that
the mass field $\bm m = (m_1, \dots, m_N)$ defines a Markov
process. Its probability distribution $p_N(\bm m, n+1)$ at time $n+1$,
which is the joint PDF of the masses in all cells, is related to that
at time $n$ by a Markov equation, which under its general form can be
written as
\begin{eqnarray}
  p_N(\bm m, n+1) &=& \int d^N\! m^{\prime}\, p_N(\bm m^{\prime}, n)\,
    P[\bm m^\prime \to \bm m] \nonumber \\ &=& \int d^N\! m^{\prime}\,
    p_N(\bm m^{\prime}, n) \int d^N \!\Omega\,\, p(\bm\Omega) \, P[\bm
    m^\prime \to \bm m | \bm\Omega],
    \label{eq:markovN}
\end{eqnarray}
where $P[\bm m^\prime \to \bm m | \bm\Omega]$ denotes the transition
probability from the field $\bm m^\prime$ to the field $\bm m$
conditioned on the realization of $\bm\Omega =
(\Omega_1,\dots,\Omega_N)$. In our case it takes the form
\begin{equation}
  P[\bm m^\prime \to \bm m | \bm\Omega] = \prod_{j=1}^{N}
  \delta[m_j-(m_j^\prime +\mu_{j-1}(n) - \mu_j(n))]\,.
\end{equation}
The variable $\mu_j$ denotes here the flux of mass between the $j$-th
and the $(j+1)$-th cell. It is a function of $\Omega_j$,
$\Omega_{j+1}$, and of the mass contained in the two cells. It can be
written as
\begin{equation}
  \mu_j(n) = \frac{\gamma}{2}\!\left[ \Omega_j(n)(1-\Omega_{j+1}(n))
    m^\prime_j(n) - \Omega_{j+1}(n)(1-\Omega_j(n))
    m^\prime_{j+1}(n)\right]\!.
\end{equation}
In the particular case we are considering, the joint probability of
the $\Omega_j$'s factorizes and we have
\begin{equation}
  p(\Omega_j) = p\,\delta(\Omega_j-1) + (1-p)\,\delta(\Omega_j)\,,
\end{equation}
so that the Markov equation (\ref{eq:markovN}) can be written in a
rather explicit and simple manner.

\begin{figure}[ht]
  \centerline{\strut\qquad
    \includegraphics[height=0.3\textwidth]{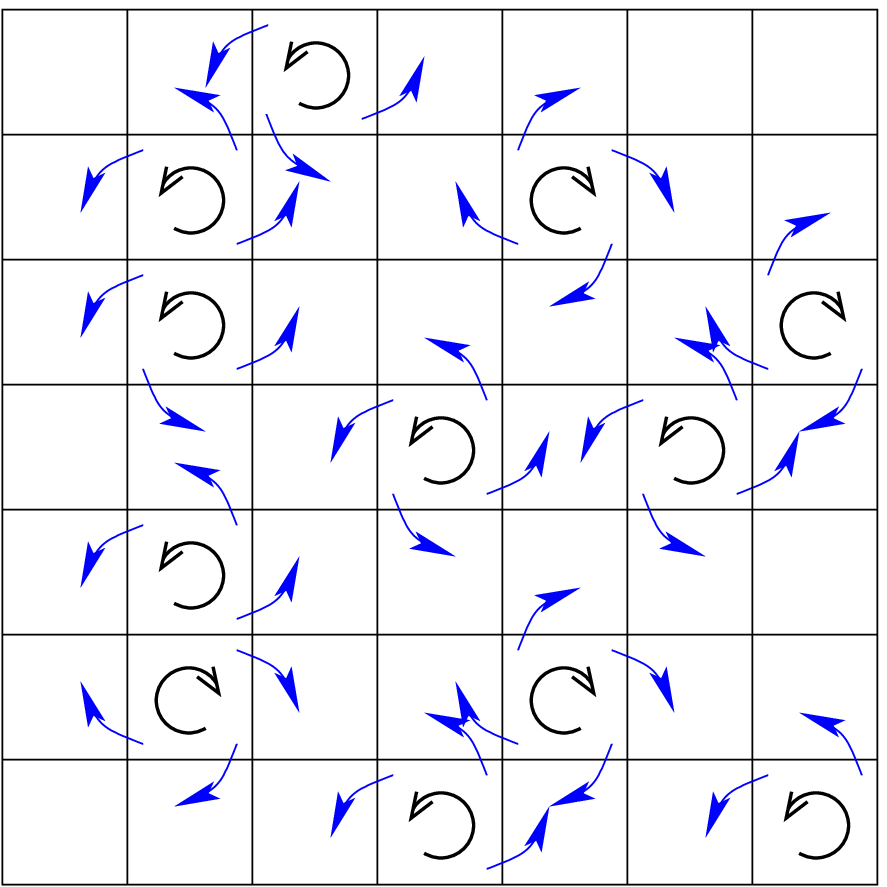} \ \
    \includegraphics[height=0.3\textwidth]{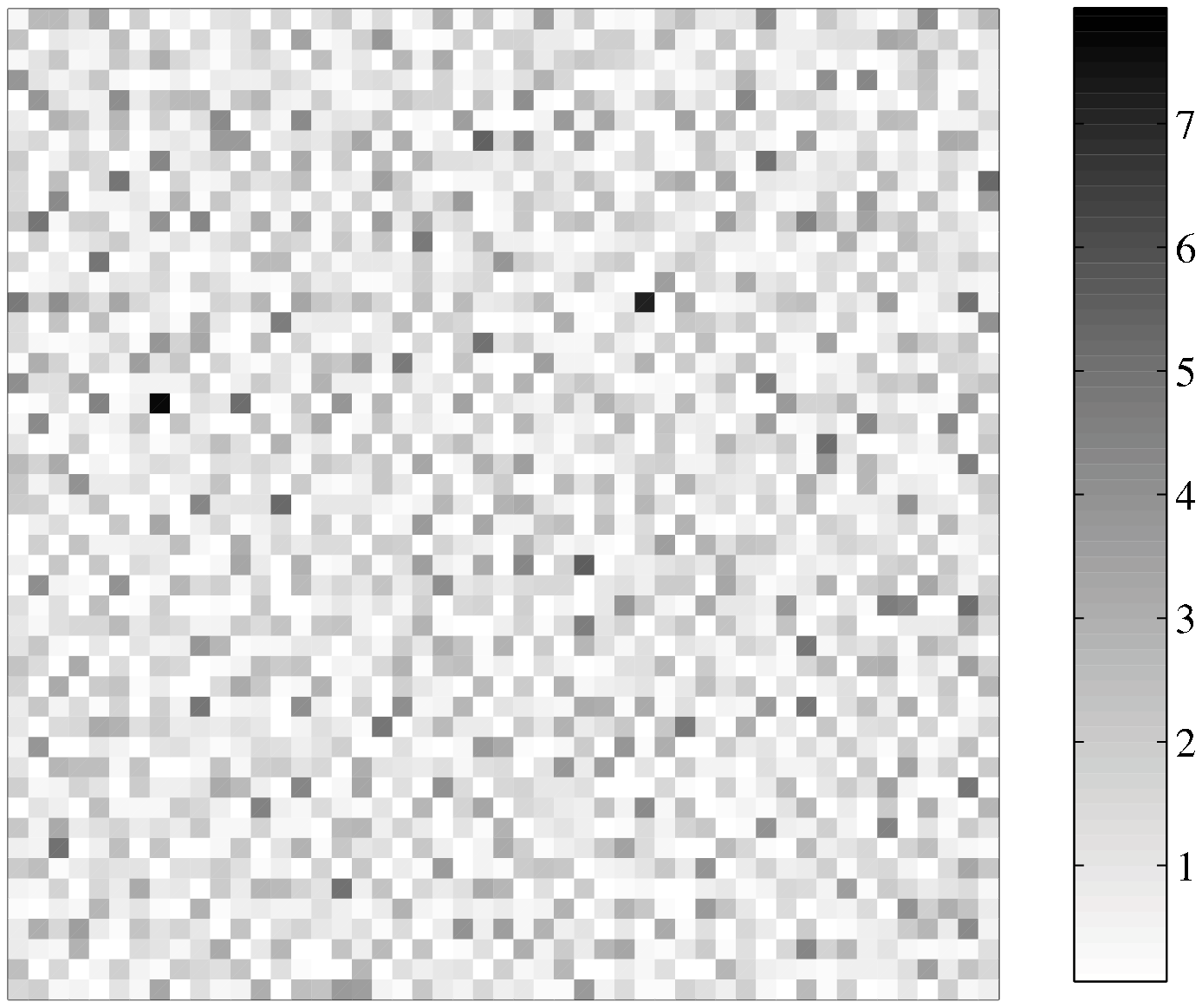}}
  \vspace{-10pt}
  \caption{Left: sketch of the ejection model in two dimensions; the
  rotating cells are designated by small eddies; the flux of mass
  (blue arrows) is from the rotating cells to those without any
  eddy. Right: snapshot of the distribution of mass in the
  statistically steady regime for a domain of $50^2$ cells with
  $p=0.75$; white squares are almost empty cells and the darker
  regions correspond to cells where the mass is concentrated.}
  \label{fig:sketch2d}
\end{figure}

The extension of the model to two dimensions is straightforward. The
mass transfer out from a rotating cell can occur to one, two, three or
four of its direct nearest neighbors (see figure~\ref{fig:sketch2d}
left). One can similarly derive a Markov equation which is similar to
(\ref{eq:markovN}) for the joint PDF $p_{N,N}(\mathbb{M},n)$ at time
$n$ of the mass configuration $\mathbb{M} = \{ m_{i,j}\}_{1\le i,j \le
N}$. The transition probability reads in that case
\begin{equation}
   P[\mathbb{M}^\prime \to \mathbb{M} | \bm\Omega] = \prod_{i=1}^{N}
  \prod_{j=1}^{N} \delta[m_{i,j}-(m_{i,j}^\prime +\mu^{(1)}_{i-1,j} -
  \mu^{(1)}_{i,j})+\mu^{(2)}_{i,j-1} - \mu^{(2)}_{i,j})]\,.
\end{equation}
where the fluxes now take the form
\begin{equation}
  \mu^{(1)}_{i,j} = \frac{\gamma}{4}\left[ \Omega_{i,j}
    (1-\Omega_{i+1,j}) \,m^\prime_{i,j}- \Omega_{i+1,j}
    (1-\Omega_{i,j}) \,m^\prime_{i+1,j}\right]
\end{equation}
and $\mu^{(2)}_{i,j}$ defined by inverting $i$ and $j$ in the
definition of $\mu^{(1)}_{i,j}$.

After a large number of time steps, a statistically steady state is
reached. The stationary distribution is obtained assuming that
$p_{N,N}(\mathbb{M},n) = p_{N,N}(\mathbb{M})$ is independent of $n$ in
the Markov equation (\ref{eq:markovN}). In this stationary state the
mass fluctuates around its mean value $1$ corresponding to a uniform
distribution; strong deviations at small masses can be qualitatively
observed (see figure~\ref{fig:sketch2d} right).

The model can be easily generalized to arbitrary space
dimension. However, as we have seen in previous section, besides its
interest from a purely theoretical point of view, the straightforward
extension to the three-dimensional case might not be relevant to
describe concentrations of inertial particles in turbulent flows.

\section{Distribution of mass}
\label{sec:pdfm}

Let us consider first the one-dimensional case in the statistically
stationary regime. After integrating (\ref{eq:markovN}), one can
express the single-point mass PDF $p_1$ in terms of the three-point
mass distribution $p_3$ at time $n$
\begin{eqnarray} p_1(m_j) &=& \int\!\!
dm^\prime_{j-1}\,dm^\prime_{j}\,dm^\prime_{j+1}\,
p_3(m^\prime_{j-1},m^\prime_{j},m^\prime_{j+1})\int\!\!
d\Omega_{j-1}\,d\Omega_{j}\,d\Omega_{j+1}\times \nonumber\\ &&
\strut\quad p(\Omega_{j-1})\,p(\Omega_{j})\,p(\Omega_{j+1})\,
\delta[m_j-(m^\prime_j + \mu_{j-1} - \mu_{j})]\,.
\label{eq:dynamics}
\end{eqnarray}
We then explicit all possible fluxes, together with their
probabilities, by considering all possible configurations of the spin
vorticity triplet $(\Omega_{j-1}, \Omega_{j}, \Omega_{j+1})$. The
results are summarized in table~\ref{tab:threecells}.  This leads to
rewrite the one-point PDF as
\begin{eqnarray}
  p_1(m) &=& \left[ p^3 + (1-p)^3 \right ]\, p_1(m) +
  \frac{2p^2\,(1-p)}{1-\gamma/2}\,
  p_1\!\left(\frac{m}{1-\gamma/2}\right) + \nonumber \\ &+&
  \frac{p\,(1-p)^2}{1-\gamma}\, p_1\!\left(\frac{m}{1-\gamma}\right)
  \nonumber + 2p\,(1-p)^2 \!\int_0^{2m/\gamma}\!\!\!\!\!\!\!\!\!\!
  dm^\prime\, p_2\!\left(m^\prime, m-\frac{\gamma}{2} m^\prime\right) +
  \nonumber\\ &+& p^2(1-p) \!\int_0^{2m/\gamma}\!\!\!\!\!\!\!\!\!\!
  dm^\prime
  \!\!\!\int_0^{2m/\gamma-m^\prime}\!\!\!\!\!\!\!\!\!\!\!\!\!\!\!\!\!\!
  dm^{\prime\prime}\,p_3\!\left(m^\prime,
  m-\frac{\gamma}{2}(m^\prime+m^{\prime\prime}),
  m^{\prime\prime}\right). \label{eq:p1form1}
\end{eqnarray}
The first term on the right-hand side comes from realizations with no
flux. The second term is ejection to one neighbor and the third to two
neighboring cells. The fourth term involving an average over the
two-cell distribution is related to events when mass is transfered
from a single neighbor to the considered cell. Finally, the last term
accounts for transfers from the two direct neighbors. Note that, in
order to write (\ref{eq:p1form1}), we made use of the fact that
$p_2(x,y)=p_2(y,x)$.
\begin{table}[ht]
  \caption{\label{tab:threecells}Enumeration of all possible
  configurations of the spin vorticity $\Omega$ in three neighboring
  cells, together with their probabilities and the associated mass
  fluxes.}
  \begin{indented}
  \item[]
    \begin{tabular}{@{}cccccc}
      \br $\Omega_{j-1}$ & $\Omega_{j}$ & $\Omega_{j+1}$ & Prob &
      $\mu_{j-1}$ & $\mu_{j}$ \\ \mr 0&0&0 & $(1-p)^3$ & 0 & 0 \\
      0&0&1 & $p\,(1-p)^2$ & 0 & $-\gamma m^\prime_{j+1}/2$ \\ 0&1&0 &
      $p\,(1-p)^2$ & $-\gamma m^\prime_{j}/2$ & $\gamma
      m^\prime_{j}/2$ \\ 0&1&1 & $p^2\,(1-p)$ & $-\gamma
      m^\prime_{j}/2$ & 0 \\ 1&0&0 & $p\,(1-p)^2$ & $\gamma
      m^\prime_{j-1}/2$ & 0\\ 1&0&1 & $p^2\,(1-p)$ & $\gamma
      m^\prime_{j-1}/2$ & $-\gamma m^\prime_{j+1}/2$\\ 1&1&0 &
      $p^2\,(1-p)$ & 0 & $\gamma m^\prime_{j}/2$\\ 1&1&1 & $p^3$ & 0 &
      0 \\ \br
    \end{tabular}
  \end{indented}
\end{table}

\begin{figure}[t]
  \centerline{\includegraphics[width=0.666\textwidth]{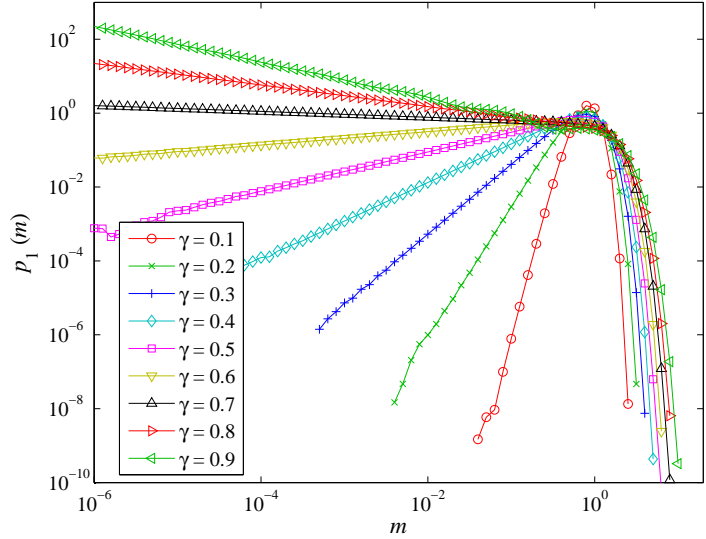}}
  \vspace{-10pt}
  \caption{Log-log plot of the one-point PDF of the mass in one
    dimension for $p=1/2$ and different values of the parameter
    $\gamma$ as labeled. The integration was done on a domain of
    $2^{16} = 65536$ cells and time average wee performed during
    $10^6$ time steps after a statistical steady state is reached.}
  \label{fig:pdfm}
\end{figure}

Numerical simulations of this one-dimensional mass transport model are
useful to grab qualitative information on $p_1$. Figure~\ref{fig:pdfm}
represents the functional form of $p_1$ in the stationary regime for
various values of the ejection rate $\gamma$ and for $p=1/2$. The
curves are surprisingly similar to measurements of the spatial
distribution of heavy particles suspended in homogeneous isotropic
turbulent flows \cite{ef94,betal06}. This gives strong evidence that,
on a qualitative level, the model we consider reproduces rather well
the main mechanisms of preferential concentration. More specifically,
a first observation is that in both settings the probability density
functions display an algebraic behavior for small masses. This implies
that the ejection from cells with vorticity one has a strong
statistical signature. A second observation is that at large masses,
the PDF decays faster than exponentially, as also observed in
realistic flows. As we will now see these two tails can be understood
analytically for the model under consideration.

\begin{figure}[h]
  \centerline{\includegraphics[width=0.666\textwidth]{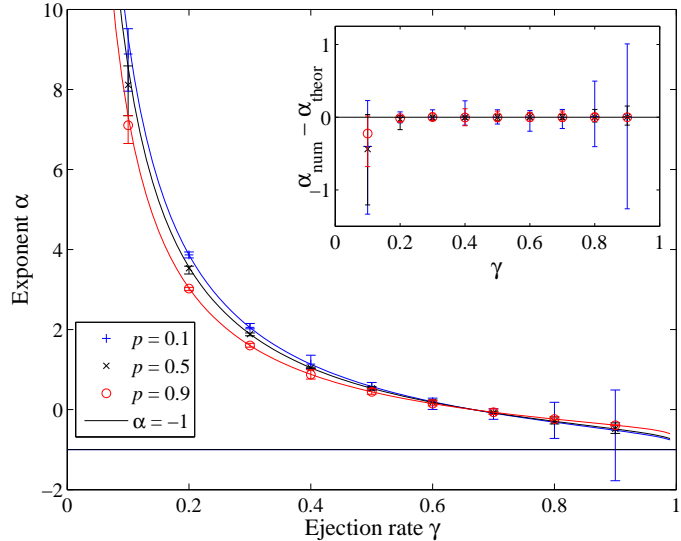}}
  \vspace{-10pt}
  \caption{Scaling exponent $\alpha$ as a function of the ejection
  rate $\gamma$ for three different values of $p$ as labeled. The
  solid lines represents the prediction given by
  (\ref{eq:alphafnT1D}); the error bars are estimated from the maximal
  deviation of the logarithmic derivative from the estimated
  value. Inset: difference between the numerical estimation and the
  value predicted by theory.}
  \label{fig:alpha_fn_T}
\end{figure}

We here first present an argument explaining why an algebraic tail is
present at small masses. For this we exhibit a lower bound of the
probability $P^<(m)$ that the mass in the given cell is less than
$m$. Namely, we have
\begin{equation}
  P^<(m) = {\rm Prob}(m_j(n)<m) \ge {\rm Prob}\,(\mathcal{A})\,,
\end{equation}
where $\mathcal{A}$ is a set of space-time realizations of $\Omega$
such that the mass in the $j$-th cell at time $n$ is smaller than
$m$. For instance we can choose the set of realizations which are
ejecting mass in the most efficient way: during a time $N$ before $n$,
the $j$-th cell has spin vorticity 1 and its two neighbors have 0. The
mass at time $n$ is related to the mass at time $n-N$ by
\begin{equation}
  m_j(n) = (1-\gamma)^N m_j(n-N)\,,\ \mbox{ that is } \ N =
  \frac{\log [m_j(n-N)/m_j(n)]}{\log (1-\gamma)}\,.
  \label{eq:Nfnm}
\end{equation}
The probability of such a realization is clearly
$p^N\,(1-p)^{2N}$. Replacing $N$ by the expression obtained in
(\ref{eq:Nfnm}), we see that
\begin{equation}
  {\rm Prob}\,(\mathcal{A}) =
  \left[\frac{m_j(n)}{m_j(n-N)}\right]^\beta \mbox{ with } \beta =
  \frac{\log [p(1-p)^2]}{\log (1-\gamma)}.
  \label{eq:pA}
\end{equation}
After averaging with respect to the initial mass $m_j(n-N)$, one
finally obtains
\begin{equation}
  P^<(m) \ge A\, m^\beta.
  \label{eq:lowbound}
\end{equation}
Hence the cumulative probability of mass cannot have a tail faster
than a power law at small arguments. It is thus reasonable to make
the ansatz that $p_1(m)$ have an algebraic tail at $m\to0$, i.e.\ that
$p_1(m) \simeq C m^\alpha$.  To obtain how the exponent $\alpha$
behaves as a function of the parameters $\gamma$ and $p$, this ansatz
is injected in the stationary version of the Markov equation
(\ref{eq:p1form1}). One expects that the small-mass behavior involves
only the terms due to ejection from a cell, namely the three first
terms in the r.h.s.\ of (\ref{eq:p1form1}), and that the terms
involving averages of the two-point and three-point PDFs give only
sub-dominant contributions. This leads to
\begin{eqnarray}
  Cm^\alpha &\approx& C\left[ p^3 + (1-p)^3 \right ] m^\alpha +
  C\,\frac{2p^2(1-p)}{1-\gamma/2}\left[\frac{m}{1-\gamma/2}\right]^\alpha
  + \nonumber\\ && + C\,\frac{p(1-p)^2}{1-\gamma}
  \left[\frac{m}{1-\gamma}\right]^\alpha\,.
  \label{eq:p1form2}
\end{eqnarray}
Equating the various constants we finally obtain that the exponent
$\alpha$ satisfies
\begin{equation}
  \frac{2p}{(1-\gamma/2)^{\alpha+1}} +
  \frac{(1-p)}{(1-\gamma)^{\alpha+1}} = 3\,. \label{eq:alphafnT1D}
\end{equation}
Note that the actual exponent $\alpha$ given by this relation is
different from the lower-bound $\beta+1$ obtained above in
(\ref{eq:pA}) and (\ref{eq:lowbound}).  However it is easily checked
that $\alpha$ approach the lower bound when $p\to 0$. As seen from
figure~\ref{fig:alpha_fn_T}, formula (\ref{eq:alphafnT1D}) is in good
agreement with numerics.
Note that the large error bars obtained for $p$ small and $\gamma$
large are due to the presence of logarithmic oscillations in the left
tail of the PDF of mass. This log periodicity is slightly visible for
$\gamma=0.9$ in figure~\ref{fig:pdfm}. It occurs when the spreading of
the distribution close to the mean value $m=1$ is much smaller than
the rate at which mass is ejected. This results in the presence of
bumps in the PDF at values of $m$ which are powers of
$1-2\gamma$. Notice that for all values of $p$, one has $\alpha\leq0$
when $\gamma\ge 2/3$. However, according to the estimate
(\ref{eq:estimgamma}), values of the ejection rate larger than $2/3$
can be attained only for large enough Kubo numbers. This is consistent
with the fact that power-law tails with a negative exponent were not
observed in the direct numerical simulations of turbulent fluid
flows~\cite{betal06} where $\Ku\approx 1$.

\begin{figure}[ht]
  \centerline{\includegraphics[width=0.666\textwidth]{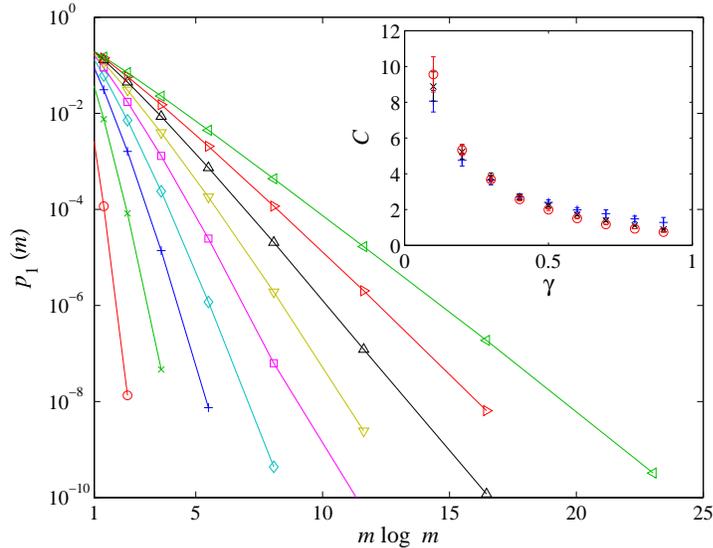}}
  \vspace{-10pt}
  \caption{Lin-log plot of the one-point PDF of the mass $m$ in one
    dimension represented as a function of $m\log m$ for $p=1/2$ and
    various values of the parameter $\gamma$ as labeled; the different
    colors and symbols are the same as those used in
    figure~\ref{fig:pdfm}. Inset: behavior of the constant $C$
    appearing in (\ref{eq:right-tail}) as a function of the ejection
    rate $\gamma$ for three different values of the fraction of space
    $p$ occupied by eddies (blue crosses: $p=0.1$, black times:
    $p=0.5$, red circles: $p=0.9$).}
  \label{fig:right_tail}
\end{figure}

It is much less easy to get from numerics the behavior of the right
tail of the mass PDF $p_1(m)$. As seen from figure~\ref{fig:pdfm},
there was no events recorded where the mass is larger than roughly ten
times its average.  We however present now an argument suggesting that
the tail is faster than exponential, and more particularly that $\log
p_1(m) \propto - m\,\log m$ when $m\gg1$. We first observe that in
order to have a large mass in a given cell, one needs to transfer to
it the mass coming form a large number $M$ of neighboring
cells. Estimating the probability of having a large mass is equivalent
to understand the probability of such a transfer.  For moving mass
from the $j$-th cell to the $(j-1)$-th cell, the best configuration is
clearly $(\Omega_{j-1},\Omega_{j},\Omega_{j+1})=(0, 1, 1)$.  After $N$
time steps with this configuration, the fraction of mass transfered is
$1-(1-\gamma/2)^{N}$.  This process is then repeated for moving mass
to the second neighbor, and so on. After order $M$ iterations, the
mass in the $M$-th neighbor is
\begin{equation}
  m = \frac{1-\left[1-(1-\gamma/2)^{N}\right]^M}{(1-\gamma/2)^{N}}\,.
  \label{eq:transfer}
\end{equation}
This means that
\begin{equation}
  M = M(m,N) = \frac{\log\left[1-m(1-\gamma/2)^{N}\right]}
  {\log\left[1-(1-\gamma/2)^{N}\right]}\,,
  \label{eq:transfer2}
\end{equation}
with the condition that $N > -(\log m)/[\log(1-\gamma/2)]$.  The
probability of this whole process of mass transfer is
\begin{equation}
  \mathcal{P} = \left[p^2(1-p)\right]^{N\,M} =
  \exp\left[\log(p^2(1-p))\,N\,M(m,N) \right]\,.
  \label{eq:probtransfer}
\end{equation}
All the processes of this type will contribute terms in the right tail
of the mass PDF. The dominant behavior is given by choosing
$N=N^{\star}$ such that $N^{\star}\,M(m,N^{\star})$ is minimal. Such a
minimum cannot be written explicitly. One however notices that, on
the one hand, if $N$ is much larger than its lower bound (i.e.\ $N \gg
-(\log m)/[\log(1-\gamma/2)]$), then $N\,M(m,N) \gg -m(\log
m)/[\log(1-\gamma/2)]$. On the other hand when $N$ is chosen of the
order of $\log m$, then $N\,M(m,N)\propto m\,\log m$. This suggests
that the minimum is attained for $N^{\star} \propto \log m$. Finally,
such estimates lead to predict that the right tail of the mass
probability density function behaves as
\begin{equation}
  p_1(m) \propto \exp\left[ -C \,m\,\log m \right]\,,
  \label{eq:right-tail}
\end{equation}
where $C$ is a positive constant that depends upon the parameters $p$
and $\gamma$. As seen in figure~\ref{fig:right_tail}, such a behavior
is confirmed by numerical experiments.

\begin{figure}[ht]
  \centerline{\includegraphics[width=0.666\textwidth]{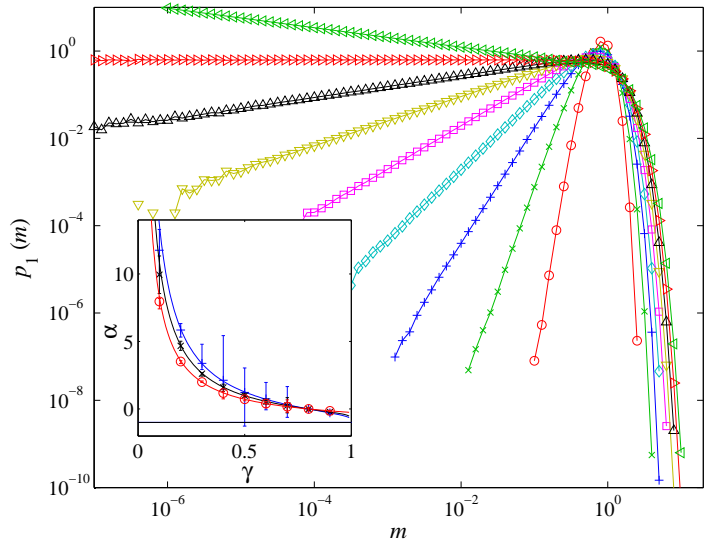}}
  \vspace{-10pt}
  \caption{Log-log plot of the one-point PDF of the mass in two
    dimensions for $p=1/2$ and various values of the parameter
    $\gamma$; the symbols and color refer to the same values of
    $\gamma$ as in figure~\ref{fig:pdfm}. The integration was done on
    a domain of $1024^2$ cells and time average were performed during
    $3\times 10^4$ time steps after a statistical steady state is
    reached. Inset: exponent $\alpha$ of the algebraic left tail as a
    function of the ejection rate $\gamma$ for three different values
    of $p$ (blue crosses: $p=0.1$, black times: $p=0.5$, red circles:
    $p=0.9$). The solid lines shows the analytic values obtained from
    (\ref{eq:alphafnT2D}).}
  \label{fig:pdfm2d}
\end{figure}

The estimations of the left and right tails of the distribution of
mass in a given cell can be extended to the two-dimensional case. The
results do not qualitatively change. The exponent $\alpha$ of the
algebraic behavior at small masses is given as a solution of
\begin{eqnarray}
  \frac{4 p^3}{(1-\gamma/4)^{\alpha+1}} &+&\frac{6
  p^2(1-p)}{(1-\gamma/2)^{\alpha+1}} +\frac{4
  p(1-p)^2}{(1-3\gamma/4)^{\alpha+1}} + \nonumber \\ &+&
  \frac{(1-p)^3}{(1-\gamma)^{\alpha+1}} =
  5(1-p+p^2)\,.\label{eq:alphafnT2D}
\end{eqnarray}
By arguments which are similar to the one-dimensional case and which
are not detailed here, one obtains also that $\log p_1(m) \propto
-m\,\log m$. Numerical experiments in two dimensions confirm these
behaviors of the mass probability distribution.  As seen from
figure~\ref{fig:pdfm2d} an algebraic behavior of the left tail of the
PDF of $m$ is observed and the value of the exponent is in good
agreement with (\ref{eq:alphafnT2D}).

\section{Coarse-grained mass distribution}
\label{sec:pdfcoarse}

\begin{figure}[ht]
  \centerline{\includegraphics[width=0.666\textwidth]{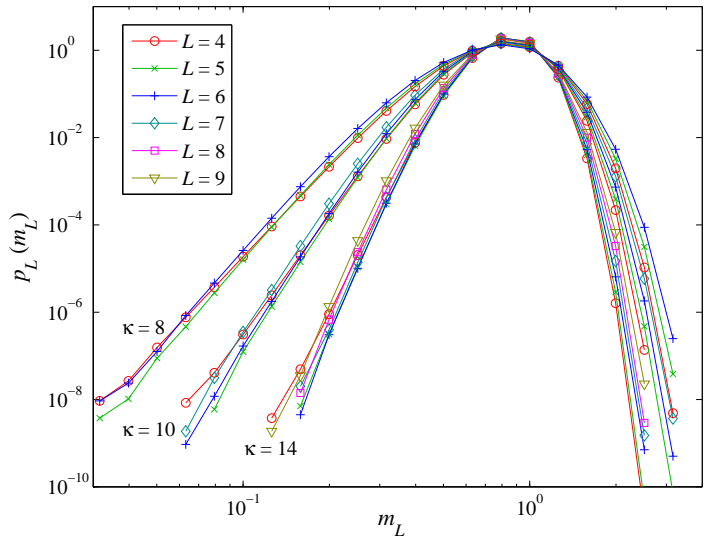}}
  \vspace{-10pt}
  \caption{Log-log plot of the PDF of the coarse-grained mass
    $\bar{m}_L$ in one dimension for $p=2/3$ and various values of
    $\gamma$ and $L$ associated to three different ratios $\kappa =
    L/\log(1-\gamma/2)$ as labeled.}
  \label{fig:pdfmbar}
\end{figure}

We investigate in this section the probability distribution of the
mass coarse-grained on a scale $L$ much larger than the box size
$\ell$, which is defined as
\begin{equation}
  \bar{m}_L = \frac{\ell}{L} \sum_{j=-K}^{K} m_j \ \mbox{ where } K =
  L/2\ell.
\label{defmbar}
\end{equation}
As seen from the numerical results presented on
figure~\ref{fig:pdfmbar}, the functional form of the PDF
$p_L(\bar{m})$ is qualitatively similar to that of the mass in a
single cell. In particular for various values of $L$ it also displays
an algebraic tail at small arguments with an exponent which depends
both on $L$ and on the parameters of the model. We here present some
heuristic arguments for the behavior of the exponent.

For this, we consider the cumulative probability $P^{<}_L(\bar{m})$ to
have $\bar{m}_L$ smaller than $\bar{m}$. We first observe that in
order to have $\bar{m}_L$ small, the mass has to be transfered from
the bulk of the coarse-grained cell to its boundaries. Assume we start
with a mass order unity in each of the $2K+1$ sub-cells.  The best
realization to transfer mass is to start with ejecting an order-unity
fraction of the mass contained in the central cell with index $j=0$ to
its two neighbors. For this the three central cells must have
vorticities $(\Omega_{-1},\Omega_0,\Omega_1) = (0, 1, 0)$,
respectively, during $N$ time steps. After that the second step
consists in transferring the mass toward the next neighbors; the best
realization is then to have during $N$ time steps
$(\Omega_{-2},\Omega_{-1}, \Omega_0, \Omega_1,\Omega_2) = (0, 1, 1, 1,
0)$. The transfer toward neighbors is repeated recursively. At the
$j$-th step, the best transfer is given by choosing
$(\Omega_{-j-1},\Omega_{-j}, \Omega_{-j+1}) = (0,1,1)$ and
$(\Omega_{j-1},\Omega_{j}, \Omega_{j+1}) = (1,1,0)$ during a time
$N$. One can easily check that for large $N$ and after repeating $K$
times this procedure, the mass which remains in the $2K+1$ cells
forming the coarse-grained cell is $\bar{m}_L \simeq
(1-\gamma/2)^{N}$. The total probability of this process is
$\left[p^4(1-p)^2\right]^{KN}$,which leads to estimate the cumulative
probability of $\bar{m}_L$ as
\begin{equation}
  P^{<}_L(\bar{m}) \propto \bar{m}^{\alpha_L}\,, \ \ \mbox{ with } \
  \alpha_L \approx \frac{1}{2}\,\frac{L}{\log(1-\gamma/2)}
  \log\left[p^4(1-p)^2\right]\,.
  \label{eq:pdfcoarse}
\end{equation} 
This approach guarantees that the probability density function
$p_L(\bar{m}) = dP^<_L/d\bar{m}$ of the coarse-grained mass $\bar{m}$
behaves as a power law at small values.  Note that only the
contribution from realizations with an optimal mass transfer is here
evaluated and the actual value of the exponent should take into
account realizations of the vorticity which may lead to a lesser mass
transfer. However we expect the estimation given by
(\ref{eq:pdfcoarse}) to hold for $L$ sufficiently large, because the
contribution from realizations with a sub-dominant mass transfer become
negligible in this limit.

As to the right tail of $p_L(\bar{m})$, one expects a behavior similar
to that obtained in the case of the one-cell mass distribution, namely
$\log p_L(\bar{m}) \propto -\bar{m}\,\log \bar{m}$ for
$\bar{m}\gg1$. Indeed, the probability of having a large mass in a
coarse-grained cell should clearly be of the same order as the
probability of having a large mass in a single cell.  This, together
with the estimates (\ref{eq:pdfcoarse}) for the exponent of the left
tail, gives a motivation for looking, at least in some asymptotics,
for possible rescaling behaviors of $p_L(\bar{m})$ as a function of
$L$ and of the ejection rate $\gamma$. For instance one can argue
whether the limits $L\to\infty$ and $\gamma\to0$ are equivalent. The
estimation (\ref{eq:pdfcoarse}) suggests that the exponent $\alpha_L$
depends only on the ratio $\kappa = L/\log(1-\gamma/2)$. Note that the
limit of small $\gamma$ should mimic that of small response time of
the heavy particles. Rescaling of the distribution of the
coarse-grained mass was observed in direct numerical simulations of
heavy particles in turbulent homogeneous isotropic
flows~\cite{betal06}.

Such a rescaling is confirmed by numerical
simulations. Figure~\ref{fig:pdfmbar} represents the PDF of the
coarse-grained mass $\bar{m}_L$ for various values of $L$ and $\gamma$
chosen such that the ratio $\kappa$ is 8, 10 or 14. While the
left-tail and the core of the distribution are clearly collapsing, the
rescaling is much less evident for the right-hand tail. Obtaining
better evidence would require much longer statistics in order to
resolve the distribution at large masses.

\section{Conclusions}
\label{sec:conclusion}

We introduced here a simple model for the dynamics of heavy inertial
particles in turbulent flow which solely accounts for their ejection
from rotating structures of the fluid velocity field.  We have shown
that this model is able to reproduce qualitatively most features of
the particle mass distribution which are observed in real turbulent
flows.  Namely the probability density of the mass in cells is shown
to behave as a power-law at small arguments and to decrease faster
than exponentially at large values. Moreover, we studied how this
distribution depends on the parameters of the model, namely the
ejection rate of particles from eddies and the fraction of space
occupied by them. Such dependence reproduce again qualitatively
observations from numerical simulations in homogeneous turbulent
flows. Finally, we have seen that coarse-graining masses on scales
larger than the cell size is asymptotically equivalent to decrease the
ejection rate related to particle inertia. This gives evidence that
there exists a scaling in the limit of large observation scale and
small response time of the particles, even if the flow has no scale
invariance.

There are several extensions that need to be investigated in order to
gain from the study of such models a more quantitative information on
the distribution of particles in real flows.  The most significant
improvement is to give a spatial structure to the fluid velocity. This
can be done by introducing a spatial correlation between the
vorticities of cells. Preliminary investigations suggest that such a
modified model could be approached by taking its continuum
limit. Another effect that may be worth taking into consideration is
random sweeping of structures by the fluid flow. We assumed that the
eddies are frozzen (and occupy the same cell) during their whole life
time. The model could be extended by adding to the dynamics random
hops between cells of these structures. Another extension could
consist in investigating in a systematic manner three-dimensional
versions of the model. As stated above, many statistical quantities
may depend on how ejection from rotating regions is implemented in
three dimensions.  Finally, it is worth mentioning again that the main
advantage of such models is to give a heuristic understanding of the
relations between the properties of the fluid velocity field and the
mass distribution of particles.  This first step is necessary in the
development of a phenomenological framework for describing the spatial
distribution of heavy particles in turbulent flows. This would allow
for using Kolmogorov 1941 dimensional arguments to understand how the
particle dynamical properties depend on scale. Moreover, such a
framework could be used to obtain refined predictions accounting for
the effect of the fluid flow intermittency and to describe the
dependence upon the Reynolds number of the spatial distribution of
particles.

\section*{Acknowledgments}

This work benefited from useful discussions with M.~Cencini and
S.~Musacchio who are warmly acknowledged.

\section*{References}
\bibliographystyle{unsrt}
\bibliography{vortices}
		
\end{document}